\documentclass[12pt,reqno]{article}
\usepackage{amsmath,amsthm,amsfonts,amssymb}
\numberwithin{equation}{section}

\newcommand{\rrangle}{{\rangle\!\rangle}}
\newcommand{\llangle}{{\langle\!\langle}}

\def \Fig#1#2#3 {
\begin{figure}
\centering
\epsfxsize=#2cm \epsfbox{#1.eps}
\caption{#3}
\label{#1}
\end{figure}
}

\newcount\figno
\def\fig#1#2#3{
\par\begingroup\parindent=0pt\leftskip=1cm\rightskip=1cm\parindent=0pt
\baselineskip=15pt
\global\advance\figno by 1
\epsfxsize=#3
\centerline{\epsfbox{#2}}
\vskip 12pt
{\bf \small Figure \the\figno:} {\small #1}\par
\endgroup\par
}
\def\L2{{\it Fun\/}\bigl(\text{\GL}\bigr)}
\def\figlabel#1{\xdef#1{\the\figno
\mbox{ }}}
\def\encadremath#1{\vbox{\hrule\hbox{\vrule\kern8pt\vbox{\kern8pt
\hbox{$\displaystyle #1$}\kern8pt}
\kern8pt\vrule}\hrule}}

\newcommand{\Pm}{\Psi^-}
\newcommand{\Pp}{\Psi^+}
\newcommand{\Ppm}{\Psi^\pm}
\newcommand{\Pmp}{\Psi^\mp}
\newcommand{\Om}{\Omega}
\newcommand{\cm}{c_-}
\newcommand{\cp}{c_+}
\newcommand{\cpm}{c_\pm}

\newcommand{\p}{\partial_p}
\newcommand{\n}{\partial_n}
\newcommand{\del}{\partial}
\newcommand{\dm}{d_-}
\newcommand{\dpp}{d_+}
\newcommand{\dpm}{d_\pm}
\newcommand{\etam}{\eta_-}
\newcommand{\etap}{\eta_+}
\newcommand{\etapm}{\eta_\pm}
\newcommand{\zm}{\zeta_-}
\newcommand{\zp}{\zeta_+}
\newcommand{\zpm}{\zeta_\pm}
\newcommand{\Ad}{\text{Ad}}
\newcommand{\ad}{\text{ad}}
\newcommand{\id}{\text{id}}

\newcommand{\str}{\text{str}}

\newcommand{\ov}{\overline}
\def\agl{{\rm $\widehat{\text{gl}}$(1$|$1)}}
\def\gl{{\rm gl(1$|$1)}}
\def\GL{{\rm GL(1$|$1)}}

\theoremstyle{break}

\newtheorem{thm}{Theorem}[section]
\newtheorem{prp}[thm]{Proposition}

\author{Thomas Creutzig$^{(1)}$, Thomas Quella$^{(2)}$ and  Volker Schomerus$^{(1)}$\\[5mm]
   $^{(1)}$ DESY Theory Group, DESY Hamburg  \\
Notkestrasse 85, D-22603 Hamburg, Germany \\
\phantom{wwwx}{\small e-mail: }{\small\tt
thomas.creutzig@desy.de, volker.schomerus@desy.de} \\[5mm]
$^{(2)}$ KdV Institute for Mathematics,
University of Amsterdam,\\
Plantage Muidergracht 24,
1018 TV Amsterdam,
The Netherlands\\
\phantom{wwwx}{\small e-mail: }{\small\tt
tquella@science.uva.nl}\\[5mm]}

\date{August 2007}

\begin{document}
\begin{titlepage}
    \title{\Huge Branes in the \GL\ WZNW-Model}
    \maketitle       \thispagestyle{empty}

\vskip1cm
\begin{abstract}
We initiate a systematic study of boundary conditions in conformal
field theories with target space supersymmetry. The WZNW model on
\GL\ is used as a prototypical example for which we find the
complete set of maximally symmetric branes. This includes a unique
brane of maximal super-dimension 2$|$2, a 2-parameter family of
branes with super-dimension 0$|$2 and an infinite set of fully
localized branes possessing a single modulus. Members of the
latter family can only exist along certain lines on the bosonic
base, much like fractional branes at orbifold singularities. Our
results establish that all essential algebraic features of
Cardy-type boundary theories carry over to the non-rational
logarithmic WZNW model on \GL.
\end{abstract}

\vspace*{-18.9cm}\noindent
{\tt {DESY 07-109}}\\
{\tt {arXiv:\,0708.0583}}

\end{titlepage}

\baselineskip=19pt
\setcounter{equation}{0}
\section{Introduction}

Field theories with target space supersymmetry have
received considerable attention lately, because of their
interesting applications in both condensed matter theory and in
string theory. This applies in particular to 2-dimensional conformal
field theories with space-time (internal) supersymmetry. They describe
critical behavior in many systems with disorder \cite{Parisi:1979ka,Efetov1983:MR708812,Bernard:1995as,Maassarani:1996jn,Zirnbauer:1999ua,Bhaseen:1999nm,Guruswamy:1999hi,Gurarie:1999yx,Ludwig:2000em,Tsvelik:2007dm}
and they provide building blocks for string theory in AdS
backgrounds
\cite{Metsaev:1998it,Berkovits:1999im,Bershadsky:1999hk,Berkovits:1999zq}.\footnote{We
should stress that in latter context,
gauge fixing the Green-Schwarz superstring leads to
non-relativistic theories which may have very different
properties from what we are about to describe.}
\smallskip

Conformal field theories with target space supersymmetry have some
properties that, for a long time, were considered rather exotic.
In fact, the correlators of such theories very often possess
logarithmic singularities on the world-sheet. In condensed matter
theory, these had been seen in various examples starting from
\cite{Rozansky:1992rx}. But it was only recently
\cite{Schomerus:2005bf} that
the appearance of logarithms in correlation functions was
understood as a rather generic consequence of internal
supersymmetry in CPT invariant local quantum field theory.
\smallskip

In many respects, logarithmic conformal field theories behave
rather differently from the well studied unitary rational models
(see, e.g., \cite{Flohr:2001zs,Gaberdiel:2001tr} and references therein).
It has proven particularly difficult to construct examples of
local logarithmic conformal field theories. Until recently, the 
only example that was fully understood was that of a triplet model 
\cite{Gaberdiel:1998ps}.
The problems may be traced back to the non-diagonalizability of the
generator $D = L_0 + \bar L_0$ of scale transformations which is
one of the characteristic features of any logarithmic conformal
field theory. Since locality implies that the generator $R = L_0 -
\bar L_0$ of rotations must be diagonalizable with integer valued
spectrum, the left and right moving sector in a logarithmic
conformal field theory must conspire in an intricate way to ensure
locality.
\smallskip

Against all these odds, recent work on WZNW models on type~I
supergroups
\cite{Schomerus:2005bf,Gotz:2006qp,Saleur:2006tf,Quella:2007hr} is now
supplying us with a large number of local logarithmic conformal field theories. This
remarkable progress is closely linked to the existence of an
action principle for these logarithmic models. The latter
furnishes valuable geometric insights in addition to efficient
computational tools. These provide an explicit solution of the
WZNW model for the supergroups \GL\ \cite{Schomerus:2005bf},
SU(2$|$1) \cite{Saleur:2006tf} and PSL(2$|$2) \cite{Gotz:2006qp} along
with powerful results and predictions for generic supergroups of
type~I \cite{Quella:2007hr}.
\smallskip

It is natural and important to extend these developments beyond
the bulk theory and to include world-sheets with boundaries. 
Systems with boundaries are highly relevant for applications
(see e.g.\ \cite{Saleur:1998hq,Saleur:2000gp,Schomerus:2002dc}
for an incomplete review of applications and many further 
references),
often more so than theories on closed surfaces. Moreover, boundary
conformal field theory also displays rather rich mathematical
structures (see e.g.\ \cite{Fuchs:2002cm,Fuchs:2003id,Fuchs:2004dz,
Fuchs:2004xi,Fjelstad:2005ua} or \cite{Schomerus:1999ug,
Alekseev:1999bs,Fredenhagen:2000ei,Maldacena:2001xj} for various
directions and further references), in particular related to 
modular properties, fusion etc.
All this is very poorly understood for general
logarithmic conformal field theories, see however 
\cite{Kogan:2000fa,Ishimoto:2001jv,Kawai:2001ur,Bredthauer:2002ct,
Bredthauer:2002xb,Pearce:2006sz} and especially
\cite{Gaberdiel:2006pp} for recent progress in specific models. 
WZNW models on supergroups present themselves as
an ideal playground to extend many of the beautiful results of
unitary rational conformal field theory to logarithmic models.
Even the simplest models are mathematically rich and physically
relevant. \smallskip

The aim of this work is to initiate a systematic study of boundary
conditions in WZNW models on supergroups based on the example of \GL.%
\footnote{Spectra of supersymmetric coset models with open boundary 
conditions were also studied previously, in particular in 
\cite{Read:2001pz,Read:2007qq}.}
Let us list the main results of this paper in more detail. Recall
that maximally symmetric boundary conditions in conformal field
theories carry two labels. The first one refers to the choice of a
gluing condition between left and right moving chiral fields. The
second label parametrizes different boundary conditions
associated with the same gluing condition. In uncompactified free
field theory, for example, the two labels correspond to the dimension
of the brane and its transverse position. The relation between these
labels and the branes' geometry becomes more intricate when the
world-sheet theory is interacting.
\smallskip
  
In the second section we shall
describe the possible ways in which we can glue left and right
movers in the \GL\ WZNW model. We shall see that there are
essentially two choices, corresponding to what we shall call {\em
untwisted} and {\em twisted} branes. Most of this work is then
devoted to the {\em untwisted branes}. We shall discuss in section~3
that all untwisted branes satisfy Dirichlet boundary conditions
for the two bosonic coordinates. Hence, they describe objects that
are fully localized in the bosonic base of the supergroup \GL. The
position of these branes is parametrized by a pair $(z_0,y_0)$ of
real numbers. For generic choices of $y_0$, the untwisted branes
extend along the two fermionic directions of \GL\ and there exists
a non-vanishing B-field. But on the lines $y_0 = 2\pi s$, for
any integer $s$, there exists an additional set of branes which
are localized in the fermionic directions as well as the bosonic
ones, i.e.\ they are truly point-like.
\smallskip
  
After a detailed study of
the branes' geometry we shall provide exact boundary states for
generic and non-generic untwisted branes on \GL\ in section~4.
There, we shall also discuss what happens when a generic brane is
moved onto one of the lines $y_0 = 2\pi s$: It turns out to split
into a pair of non-generic branes with a transverse separation
that is proportional to the level of the WZNW model. Section~5
contains a detailed discussion of the relation between our
findings for boundary conditions in a local logarithmic conformal
field theory and the usual Cardy case of unitary rational models
\cite{Cardy:1989ir}.
We shall see that in both cases branes are parametrized by
irreducible representations of the current algebra. Furthermore,
the spectra between any two branes can be determined by fusion.
Similar results for the $p=2$ triplet model have been obtained 
in \cite{Gaberdiel:2006pp}. In the case of \GL\ WZNW model we will 
establish that most of the boundary spectra are not fully 
reducible. This applies in particular to the spectrum of 
boundary operators on a single generic brane. Section~6 is 
devoted to a brief study of twisted branes on \GL. We shall 
find that these satisfy Neumann boundary conditions in the 
bosonic coordinates.

\section{Gluing Conditions for $\mathbf{\widehat{gl}(1|1)}$ Symmetric Branes}

Branes on supergroups come in different families or types. They
are characterized by the way in which left and right moving chiral
fields are glued along the boundary (see e.g.\ \cite{Recknagel:1997sb}).  
Mathematically, the various
possible gluing conditions correspond to automorphisms of the
chiral symmetry. If two gluing automorphisms differ by an inner
automorphism, the associated branes are related to each other by
simple translation on the target space.

The chiral symmetry of the \GL\  WZNW model is a \agl\  current
superalgebra. Its metric preserving automorphisms 
will be classified in the first
subsection up to the possible composition with an inner
automophism. In addition to the trivial automorphism we shall find
one non-trivial outer automorphism $\Omega$. Some general facts
about the associated gluing conditions for supercurrents and their
geometrical interpretation are collected in the second subsection.

\subsection{Automorphisms of the $\mathbf{\widehat{gl}(1|1)}$ current superalgebra}

In this subsection, we determine the relevant gluing automorphisms
$\Omega$ for branes in the \GL\  WZNW model. An automorphism of
the \agl\  current superalgebra is admissible as a gluing
automorphism if it acts trivially on the Virasoro Sugawara field
$T$. When restricted to the zero mode algebra, any such
automorphisms $\Omega$ gives rise to an automorphism $\omega$ of
the underlying finite dimensional Lie superalgebra \gl.
If $\Omega$ leaves $T$ invariant, the corresponding automorphism
$\omega$ acts trivially on the associated quadratic Casimir element
$C$ of \gl. Our first goal is therefore to classify all automorphisms
$\omega$ of \gl\ with the additional property that $\omega(C) = C$.
\smallskip

The Lie superalgebra \gl\ is generated by two bosonic elements
$E,N$ and two fermionic elements $\Psi^\pm$, subject to the
relations
\begin{equation}
    [N,\Psi^\pm]\ =\ \pm\Psi^\pm \ \ ,
     \ \ \{\Psi^-,\Psi^+\}\ =\ E \ \ .
\end{equation}
In addition, the element $E$ is central, i.e.\ it commutes with
all other elements of \gl. The relevant quadratic Casimir element
$C$ of \gl\ is given by
\begin{equation}
    C\ =\ (2N-1)E+2\Pm\Pp + \frac{1}{k} E^2\ \ .
\end{equation}
Since $E$ is central, one has the freedom of adding a
quadratic polynomial in $E$. The choice we have made here is the
one that corresponds to the Virasoro Sugawara field of the \agl\
current superalgebra at level $k$ that has been used in
\cite{Schomerus:2005bf}. In this context the subleading term in $k$
should be thought as a quantum renormalization. Adding additional
contributions in $E^2$ does not change the qualitative features of the
model. \smallskip

A straightforward calculation shows, that the Casimir preserving
automorphisms of \gl\ come in two families,
\begin{alignat}{7} \label{eq:Om}
        \omega^{(0)}_\alpha(E) & = & E \ \ , \ \
        \omega^{(0)}_\alpha(N) & = & N \ \ , \ \
        \omega^{(0)}_\alpha(\Ppm) & = & e^{\pm i\alpha} \Ppm \
        &&\\[2mm]
        \omega^{(1)}_\alpha(E) & = & \ - E \ \ , \ \
        \omega^{(1)}_\alpha(N) & = & \ - N \ \ , \ \
        \omega^{(1)}_\alpha(\Ppm) & = & \ \pm e^{\pm i\alpha}\Pmp \
         &.&
\end{alignat}
With $E$ being central, the only non-trivial bosonic inner
automorphisms $\Ad_\alpha$ are provided by conjugation with $\exp
(i\alpha N)$. Looking back onto the eqs.\ \eqref{eq:Om}, we
observe that $\omega^{(0)}_\alpha = \Ad_\alpha$, i.e.\ the
automorphisms $\omega^{(0)}_\alpha$ are all inner. Furthermore, any
two members of the second family $\omega^{(1)}_\alpha$ are related by
conjugation with some $\exp (i \alpha N)$. Hence, it suffices to
consider one representative $\omega = \omega^{(1)}_{\alpha=0}$. We
conclude that, up to composition with inner automorphisms, there
exist two admissible automorphisms of \gl, namely the trivial
automorphism $\omega^{(0)} = \id$ and the non-trivial $\omega =
\omega^{(1)}_0$. Note that the latter squares to an inner 
automorphism. 
\smallskip

Let us now show that both automorphisms lift to admissible
automorphisms of the current superalgebra \agl. This current
algebra is generated by the modes of the chiral fields $E(z),
N(z)$ and $\Ppm(z)$ with relations,
\begin{equation} \label{eq:kmcomm}
[E_n,N_m]\ =\ -km\delta_{n+m} \ , \ \ [N_n,\Ppm_m]\ =\
\pm\Ppm_{n+m}\ , \ \ \{\Pm_n,\Pp_m\}\ =\ E_{n+m}+km\delta_{n+m}\ .
\end{equation}
All other \mbox{(anti-)}commutators vanish and the number $k$ is known as
the level of \agl. The action of $\omega^{(0)} = \id$ on \gl\ lifts to
the trivial automorphism $\Omega^{(0)} = \id$ on \agl. In case of
$\omega^{(1)}$, its properties guarantee that
$$ \Omega(E_n) \ =\  - E_n \ \ , \ \
   \Omega(N_n) \ = \ - N_n \ \ , \ \
   \Omega(\Ppm_n) \ = \ \pm \Pmp_n
$$
is consistent with the level dependent terms in eqs.\ %
\eqref{eq:kmcomm}. Furthermore, the modes of the stress energy
tensor take the form \cite{Rozansky:1992rx}
\begin{eqnarray*}
 L_n & = &  \frac{1}{2k}\  (2 N_n E_0- E_n +
       2 \Psi^-_n \Psi^+_0 +\frac{1}{k} E_n E_0) \\[2mm]
        & & +  \frac{1}{k} \sum_{m > 0} \, (E_{n-m} N_m +
    N_{n-m} E_m + \Psi^-_{n-m} \Psi^+_{m} - \Psi^+_{n-m} \Psi^-_m
    + \frac{1}{k} E_{n-m} E_m)
\end{eqnarray*}
It is easy to check that the $L_n$ are indeed invariant under the
action of $\Omega$. Consequently we have found two classes of
automorphisms of \agl\ that are admissible as gluing
automorphisms.
 
\subsection{Types of boundary conditions}

Let us consider a WZNW model on the upper half of the complex plane.
Boundary conditions along the boundary at $z= \bar z$ preserve
conformal invariance of the model if  and only if the two chiral
components of the stress energy tensor $T$ agree all along the
boundary, i.e.\
\begin{equation}
    T(z)\ =\ \ov{T}(\bar{z}) \ \ \text{for} \ z = \bar{z} \ .
\end{equation}
In any WZNW model, the stress energy tensor $T$ is constructed out
of the chiral currents. A boundary condition is said to be
maximally symmetric if left and right moving currents can be
identified along the boundary, up to the action of an automorphism
$\Omega$,
\begin{equation} \label{eq:glue}
    J^a(z)\ =\ \Omega\bigl(\bar{J^a}(\bar{z})\bigr) \ \ \text{for} \
                                      z = \bar{z}
    \ .
\end{equation}
where $J^a = E,N,\Ppm$ when we deal with the \GL\ model. For
$\Omega$ we can insert any of the automorphisms we have discussed
in the previous subsection.
\smallskip

It will be convenient to rewrite the gluing conditions
\eqref{eq:glue} in terms of those fields that appear in the action
of the \GL\  WZNW model. In principle, there exist various choices
that come with different parametrizations of the supergroup \GL.
One possible set of coordinate fields is introduced through
\begin{equation}\label{eq:par1}
    g\ =\ e^{i\cm\Pm}\, e^{iXE+iYN}\, e^{i\cp\Pp}\ \ .
\end{equation}
The fields $X$ and $Y$ are bosonic while $\cpm$ are fermionic.
Let us also recall that the \mbox{(anti-)}holomorphic currents of the
WZNW model are given by
$$
   J(z)\ =\ -k\del g g^{-1} \ \ \ {\text{and}} \ \
   \bar{J}(\bar{z}) \ =\ k g^{-1}\bar{\del} g \ \ .
$$
Inserting our specific choice of the paramerization
\eqref{eq:par1}, the currents take the following form
\begin{equation} \label{eq:J}
    \bar{J}\ = \ kie^{iY}\bar{\del}\cm\Pm + k\bigl(i\bar{\del}X -
     (\bar{\del}\cm)\cp e^{iY}\bigr) E + ki\bar{\del}Y N +
     k(i\bar{\del} \cp- \cp \bar{\del} Y)\Pp
\end{equation}
and
\begin{equation} \label{eq:bJ}
    J \ = \ -k(i\del\cm-\cm\del Y)\Pm -k\bigl(i\del X - \cm (\del \cp) e^{iY}\bigr)
    E -k i\del Y N -k ie^{iY}\del\cp\Pp.
\end{equation}
The various components of these Lie superalgebra valued
\mbox{(anti-)}holomorphic currents can be projected out with the help of
the super-trace
\begin{equation}
    \str(NE)\ =\ \str(\Pp\Pm)\ =\ -1\ .
\end{equation}
We conclude that $E(z) = \str(J(z)E) = ki \partial Y$ and similar
expressions hold for the other three holomorphic currents
and their \mbox{anti-}holomorphic counterparts.
\smallskip

Let us briefly recall how to extract the branes' geometry from the
gluing conditions. Locally, the action of a WZNW model on any
(super-)group looks as follows
\begin{equation}
S(X)\ \sim \ \int_\Sigma d^2z (g_{\mu\nu}+B_{\mu\nu})\del X^\mu
  \bar{\del}X^\nu.
\end{equation}
with a (graded) antisymmetric 2-form potential $B$ of the WZ
3-form $H = dB$ and a (graded) symmetric metric $g$. Vanishing of
the boundary  contributions to the variation leaves us with two
choices: We can either impose Dirichlet boundary conditions $\p
X^\mu=0$ or require that
\begin{equation} \label{eq:NwBbc}
      g_{\mu\nu}\,\n X^\mu(z,\bar{z}) \ =\
       iB_{\mu\nu}\,\p X^\mu(z,\bar{z}) \ \ \text{for} \ z = \bar{z}\ .
\end{equation}
In general, some combination of these two possibilities occurs.
The gluing conditions \eqref{eq:glue} for our currents
\eqref{eq:J} and \eqref{eq:bJ} can always be brought into standard
form by splitting the derivatives $\partial$ and $\bar \partial$
into $\p$ and $\n$. Following the reasoning that was first
proposed in \cite{Alekseev:1998mc} for bosonic WZNW models (see also
\cite{Felder:1999ka} for a different approach), one
may show that maximally symmetric branes on super-groups are
localized along $\omega$ twisted super-conjugacy classes
\begin{equation}
    C^\omega(b)\ =\ \bigl\{\omega(g)bg^{-1}\ \bigl| \ g \ \text{in} \ G \bigr\}
\end{equation}
where $b$ can be any element of the bosonic subgroup and $\omega$
is now regarded as an automorphism of the supergroup rather than
its Lie superalgebra. For the \GL\ WZNW model, a more detailed
derivation of this statement along with an explicit description of
the resulting brane geometries will be given below.

\section{Untwisted Branes: Geometry and Particle limit}

This section is devoted to the geometry of branes associated with
the trivial gluing automorphism. We shall show that such branes
are localized at a point $(x_0,y_0)$ on the bosonic base of \GL.
For generic choices $y_0$, they stretch out along the fermionic
directions, i.e.\ the fermionic fields obey Neumann type boundary
conditions. When $y_0 = 2\pi s, s \in \mathbb{Z}$, on the other
hand, the corresponding branes are point-like. These geometric
insights from the first part of the section are then used in the
second part to study branes in the particle limit in which the
level $k$ is sent to infinity. Most importantly, we shall
provide minisuperspace analogues of the boundary states for
both generic and non-generic untwisted branes, see eqs.\ %
\eqref{eq:MSSbst1} and \eqref{eq:MSSbst2}, respectively.

\subsection{Geometric interpretation of untwisted branes}

In the previous section we have made a number of general statements
concerning the geometry of maximally symmetric branes on (super-)group
target spaces. Here, we want to step back a bit and work out the
precise form of the boundary conditions for coordinate fields. We
shall continue to use the specific parametrization \eqref{eq:par1}
of \GL. Insertion of our explicit formulas \eqref{eq:J} and
\eqref{eq:bJ} for left and right moving currents into the gluing
condition \eqref{eq:glue} with $\Omega= \mathbb{I}$ gives
\begin{equation} \label{eq:BUG}
    \begin{split}
        &\p Y \ = \ 0 \qquad , \qquad \p Z\  = \ 0 \qquad ,
         \qquad \text{for} \ z\, =\, \bar{z} \ , \\[2mm]
        &\text{where} \qquad Z=X+i\cm\cp(e^{-iY}-1)^{-1} \  \\
        \end{split}
\end{equation}
and $\partial_p$ denotes the derivative along the boundary. In
other words, both bosonic fields $Y$ and $Z$ satisfy Dirichlet
boundary conditions. Untwisted branes in the \GL\ WZNW model are
therefore parameterized by the constant values $(y_0,z_0)$ the two
bosonic fields $Y,Z$ assume along the boundary. For the two basic
fermionic fields we obtain similarly
\begin{equation} \label{eq:FUG}
    \begin{split}
        &\pm 2 \sin^2(Y/2)\n \dpm \ = \ \sin(Y)\,\p \dpm \qquad ,
         \qquad \text{for} \ z\, =\, \bar{z}\ , \\[2mm]
        &\text{where} \qquad\dpm\ =\ \cpm e^{iY/2}\sin^{-1}(Y/2)/2i\
        .\\
         \end{split}
\end{equation}
Thereby, the fermionic directions are seen to satisfy Neumann
boundary conditions with a constant B-field whose strength
depends on the position of the brane along the bosonic base. We
shall provide explicit formulas below. For the moment let us point
out that the condition \eqref{eq:FUG} degenerates whenever the
value $y_0$ of the bosonic field $Y$ on the boundary approaches an
integer multiple of $2\pi$. In fact, when $y_0 = 2\pi s, s \in
\mathbb{Z}$ we obtain Dirichlet boundary conditions in all
directions, bosonic and fermionic ones,
\begin{equation}
    \p Y\ =\ \p Z=\p \dpm \ =\ 0 \ \ \text{for} \ z\, =\, \bar{z}.
\end{equation}
In the following, we shall refer to the branes with parameters
$(z_0,y_0 \neq 2\pi s)$ as {\em generic (untwisted)
branes}. These branes are localized at the point $(z_0,y_0)$ of
the bosonic base and they stretch out along the fermionic
directions. A localization at points $(z_0,2\pi s),
s \in \mathbb{Z}$, implies Dirichlet boundary conditions for
the fermionic fields. We shall refer to the corresponding
branes as {\em non-generic (untwisted) branes}.
\smallskip

We have seen in the description of our gluing conditions that it
was advantageous to introduce fields $Z$ and $d_\pm$ instead of
$X$ and $\cpm$. They correspond to a new choice of coordinates on
the supergroup \GL\
\begin{equation}\label{eq:par}
    g\ =\ e^{i\cm\Pm}e^{ixE+iyN}e^{i\cp\Pp}\ =\
     e^{i\dm\Pm}e^{-i\dpp\Pp}e^{izE+iyN}e^{i\dpp\Pp}e^{-i\dm\Pm}
\end{equation}
that is particularly adapted to the description of untwisted branes.
In fact, we recall from our general discussion that untwisted branes
are localized along conjugacy classes. It is therefore natural to
introduce a parametrization in which supergroup elements $g$ are
obtained by conjugating bosonic elements $g_0 = \exp (iz_0E +iy_0N)$
with exponentials of fermionic generators. From equation \eqref{eq:par}
it is also easy to read off that conjugacy classes containing a bosonic
group element $g_0$ contain two fermionic directions as long as $y_0
\neq 2\pi s$. In case $y_0 = 2\pi s$, conjugation of $g_0$ with the
fermionic factors is a trivial operation and hence the conjugacy
classes consist of points only.
\smallskip

It is instructive to work out the form of the background metric
and B-field in our new coordinates. To this end, let us recall
that
\begin{equation} \label{eq:sold}
ds^2\ =\ \str\bigl((g^{-1}dg)^2\bigr)\ =\ 2dxdy-2e^{iy}d\etam d\etap\ \ .
\end{equation}
Here, the super-coordinates $x,y,\etapm$ correspond to our
coordinate fields $X,Y,c_\pm$. Similarly, the Wess-Zumino
3-form on the supergroup \GL\ is given by
\begin{equation}\label{Hold}
H\ =\ \frac{2}{3}\; \str(g^{-1}dg)^{\wedge 3}\ =\
2ie^{iy}d\etam\wedge d\etap\wedge dy\ \ .
\end{equation}
After the appropriate change of coordinates from $(x,y,\etapm)$
to $(z,y,\zpm)$, the metric reads
\begin{equation} \label{eq:snew}
    ds^2\ =\ 2dzdy+8\sin^2(y/2)d\zm d\zp
\end{equation}
and the $H$ field becomes
\begin{equation} \label{eq:Hnew}
H \ = \ 4 i \bigl(\cos(y)-1\bigr) d\zm\wedge d\zp \wedge d
y \ \ .
\end{equation}
It is easy to check that $H = dB$ possesses a 2-form potential
$B$ given by
\begin{equation} \label{Bnew}
    B\ =\ 4i\sin(y)\,d\zm\wedge d\zp\ + 2i\zp d\zm\wedge dy -
         2i\zm d\zp \wedge dy \ \ .
\end{equation}
Upon pull back to the untwisted branes we can set $dy=0$ and the
B-field becomes
\begin{equation}
 \pi^\ast_{\text{brane}}\, B\ = \ 4i\sin(y)\,d\zm\wedge d\zp\ \ .
\end{equation}
This expression together with our formula \eqref{eq:snew} for the
metric allow to recast the boundary conditions \eqref{eq:FUG} for
the fermionic fields in theories with generic untwisted boundary
conditions in the familiar form \eqref{eq:NwBbc}.

\subsection{Boundary states in the minisuperspace theory}

As in the analysis of the bulk \GL\ model \cite{Schomerus:2005bf}
it is very instructive to study the properties of untwisted branes
in the so-called particle or minisuperspace limit. Thereby we
obtain predictions for several field theory quantities in the
limit where the level $k$ tends to infinity. Our first aim is
to present formulas for the minisuperspace analogue of Ishibashi
states. Using our insights from the previous subsection we shall
then propose candidate boundary states for the particle limit
and expand them in terms of Ishibashi states.
\smallskip

Let us begin by recalling a few basic facts about the supergroup
\GL\ or rather the space of functions $\L2$ it determines, see
\cite{Schomerus:2005bf}. The latter is spanned by the elements
\begin{equation} \label{eq:basis}
e_0(e,n)\ =\ e^{iex+iny}\ , \ \ e_\pm(e,n)\ =\ \etapm e_0(e,n)\, \
\ \  e_2(e,n)\ =\ \etam\etap e_0(e,n) \ .
\end{equation}
where the coordinates are the same as in the previous subsection.
Right and left invariant vector fields take the following form
\begin{equation}
R_E\ =\ i\del_x \ , \ \ R_N\ =\ i\del_y+\etam\del_- \ , \ \ R_+\
=\ -e^{-iy}\del_+-i\etam\del_x \ , \ \ R_-\ =\ -\del_-\ ,
\end{equation}
and
\begin{equation}
L_E\ =\ -i\del_x \ , \ \ L_N\ =\ -i\del_y-\etap\del_+ \ , \ \ L_-\
\ =\ e^{-iy}\del_- -i\etap\del_x \ , \ \ L_+\ =\ \del_+\ ,
\end{equation}
These vector fields generate two \mbox{(anti-)}commuting copies of the
underlying Lie superalgebra \gl. For the reader's convenience we
also wish to reproduce the invariant Haar measure on \GL,
\begin{equation} d\mu \ = \ e^{-iy}dxdyd\etap d\etam\ \ .
\end{equation}
The decomposition of $\L2$ with respect to both left and right
regular action was analyzed in \cite{Schomerus:2005bf}. Here, we
are most interested in properties of the adjoint action $\ad_X =
R_X + L_X$ since it is this combination of the symmetry generators
that is preserved by the untwisted D-branes.
\smallskip

Our first aim is to construct a canonical basis in the space of
(co-)invariants. By definition, a (co-)invariant $|\psi\rrangle$
($\llangle \psi |$) is a state (linear functional) satisfying
\begin{equation} \label{eq:MSSinv}
 \ad_X |\psi \rrangle \ = \ (R_X + L_X) |\psi\rrangle \ = \ 0 \ \ \ , \ \
\ \llangle \psi | \,\ad_X \ = \ \llangle \psi| (R_X + L_X) \ = \ 0 \
\ .
\end{equation}
These two linear conditions resemble the so-called
Ishibashi conditions in boundary conformal field theory. In the
minisuperspace theory, it is easy to describe the space of
solutions. One may check by a short computation that a generic
invariant takes the form
\begin{equation}
|e,n\rrangle_0 \ = \ \frac{1}{2\pi \sqrt e} \bigl( e_0(e,n)- e_0(e,n-1)
+ee_2(e,n)\bigr)  \ \ .
\end{equation}
The pre-factor $1/2\pi \sqrt e$
is determined by a normalization condition to be spelled out
below. We note that the function $|e,n\rrangle_0$ is obtained by
taking the super-trace of supergroup elements in the typical
representation $\langle e,n\rangle$.\footnote{Our conventions for
the representation theory of \gl\ are the same as in
\cite{Gotz:2005jz}. In particular, $\langle e,n\rangle$ denotes a
2-dimensional graded representation of \gl. Let us agree to
consider the state with smaller $N$-eigenvalue as even (bosonic).
The same representation with opposite grading shall receive an
additional prime, i.e.\ it is denoted by $\langle e,n\rangle'$.}
To each of the invariants $|e,n\rrangle_0$ we can assign a
co-invariant $_0\llangle e,n|:\L2 \rightarrow \mathbb{C}$ through
\begin{equation}
\ _0\llangle e,n| \ = \ \int d\mu \,\frac{1}{2 \pi \sqrt e} \bigl(
e_0(-e,-n+1)- e_0(-e,-n) -ee_2(-e,-n+1)\bigr)\ .
\end{equation}
Our normalization of both $|e,n\rrangle_0$ and the dual invariant
$_0\llangle e,n|$ ensures that
$$\ _0\llangle e,n| (-1)^F u_1^{\frac12(L_E-R_E)} u_2^{\frac12 (L_N-R_N)}
 |e',n'\rrangle_0 \ = \ \delta(n'-n) \, \delta(e'-e)
 \, \chi_{\langle e,n\rangle} (u_1,u_2)$$
where $ \chi_{\langle e,n\rangle} (u_1,u_2) =
u_1^e\left(u_2^{n-1}- u_2^{n}\right)$ is the super-character of
the typical representation $\langle e,n\rangle$ of \gl. If we
re-scale the invariants $|e,n\rrangle_0$ and then send $e$ to zero
we obtain another family of invariants,
\begin{equation}
|0,n\rrangle_0 \ := \  \lim_{e\rightarrow 0} \sqrt{e}\,
|e,n\rrangle_0 \ = \ e_0(0,n) - e_0(0,n-1) \ \ .
\end{equation}
Similarly, we define the dual $\ _0\llangle 0,n|$ as a limit of $\
_0\llangle -e,-n+1| \sqrt{e}$. By construction, the states
$|0,n\rrangle_0$ and the associated linear forms possess
vanishing overlap with each other and with the states
$|e,n\rrangle_0$,
\begin{equation}
\ _0\llangle 0, n| u_1^{\frac12(L_E-R_E)} u_2^{\frac12 (L_N-R_N)}
 |e',n'\rrangle_0 \ = \ 0
\end{equation}
for all $e'$, including $e'=0$. This does certainly not imply that
$\ _0\llangle 0,n|$ acts trivially on the space of functions.
\smallskip

It is easy to see that the functions $|0,n\rrangle_0$ do not yet
span the space of invariants. What we are missing is a family of
additional states $|n\rrangle_0$  which is given by
$$ |n\rrangle _0 \ = \ \frac{1}{2\pi}\,  e_0(0,n) \ \ \ \mbox{ for } \ \ \
   n \ \in \ [0,1[ \ \ . $$
The corresponding dual co-invariants are given by the prescription
\begin{equation} \label{eq:nd}
\ _0\llangle n| \ = \ \frac{1}{2\pi} \, \int d\mu \, 
  \sum_{m \in \mathbb{Z}} e_2(0,-n+m+1) \ \ . 
\end{equation}
Our normalization ensures that
\begin{equation}
\ _0\llangle n| (-1)^F u_1^{\frac12(L_E-R_E)} u_2^{\frac12
(L_N-R_N)} |n'\rrangle_0 \ = \ \delta(0) \, \delta(n'-n)\, 
\chi_{\langle n\rangle} (u_1,u_2)
\end{equation}
where $\chi_{\langle n \rangle}(u_1,u_2) = u_2^n$. The divergent 
factor $\delta(0)$ stems from the infinite volume of our target 
space and it could absorbed into the normalization of the 
Ishibashi state. Let us observe that the co-invariants $ _0 
\llangle n|$ may be obtained by a limiting procedure from 
$\ _0\llangle e,n|$,
\begin{equation} \label{eq:nlimit}
\ _0 \llangle n | \ = \ - \lim_{e\rightarrow 0}\
\frac{1}{\sqrt{e}} \ \sum_m \, _0\llangle e,n+m|\ \ .
\end{equation}
A similar construction can be performed with the Ishibashi states
$|e,n\rrangle_0$ to give the formal invariants $\sum_m e_2(0,n+m)$.
They are formally dual to co-invariants given by $\int d\mu
e_0(0,-n+1)$. In our discussion, and in particular when we wrote
eq.\ \eqref{eq:nd}, we have implicitly equipped $\L2$ with a
topology that excludes to consider $\sum_m e_2(0,n+m)$ as
a true function. While the dual functional $\int d\mu
e_0(0,-n+1)$ does not suffer from any such problem, it so happens
not to appear in the construction of boundary states. This is why
we do not bother giving it a proper name.
\medskip

It is our aim now to determine the coupling of bulk modes to
branes in the minisuperspace limit. In the particle limit, the
bulk 1-point functions are linear functionals $f \mapsto \langle f
\rangle$ on the space $\L2$ of functions such that $\langle \ad_X
f \rangle = 0$, i.e.\ they are co-invariants. The first family of
co-invariants we shall describe corresponds to branes in generic
positions $(z_0,y_0)$. Since these are localized at a point
$(z_0,y_0)$ on the bosonic base and delocalized along the
fermionic directions, their density is given by
\begin{equation}
  \begin{split}
\rho_{(z_0,y_0)} & =\ -2i \sin(y_0/2)\, \delta(y-y_0)
\, \delta(z-z_0) \\[2mm]
& =\ - 2 i \sin(y_0/2)\,  \delta(y-y_0) \, \delta\bigl(x - i
\etam\etap (1-e^{-iy})^{-1} - z_0\bigr) \ \ .
  \end{split}
\end{equation}
The constant prefactor $-2i\sin(y_0/2)$ was chosen simply to 
match the normalization of our boundary states below.  
Obviously, the density $\rho_{(z_0,y_0)}$ is invariant under the
adjoint action. It gives rise to a family of co-invariants through
the prescription
\begin{equation} \label{eq:coinv}
 f \mapsto \langle f \rangle_{\rho}  \ :=\
   \int d\mu \, \rho(x,y,\etapm) \, f(x,y,\etapm)\ \ .
\end{equation}
Geometrically, the integral computes the strength of the coupling
of a bulk mode $f$ to a brane with mass density $\rho$. It is not
difficult to check that our functional $\langle \cdot
\rangle_{(z_0,y_0)}$ admits an expansion in terms of dual
Ishibashi states as follows,
\begin{equation} \label{eq:MSSbst1}
\begin{split}
\langle\  \cdot\  \rangle_{(z_0,y_0)} & \equiv \ _0\langle z_0, y_0| \ = \
   \int dedn \, \sqrt{e} \,
e^{i(n-1/2) y_0 +iz_0e} \ _0 \llangle e,n|\\[2mm]
& \hspace*{-2cm} = \ \int_{e\neq 0} dedn\,\sqrt{e}\,e^{i(n-1/2) y_0
+iz_0e} \ _0 \llangle e,n| + \int dn \, e^{i(n-1/2) y_0} \
_0\llangle 0,n|\ \ .
\end{split}
\end{equation}
In the second line of this formula we have separated typical and
atypical contributions to the boundary state. Considering that the
state $_0\llangle 0,n|$ is obtained through the limiting procedure
$_0\llangle 0,n| = \lim_{e\rightarrow 0} \sqrt {e} \
_0\llangle e,n|$, the second term is the natural continuation of
the first. In this sense, we may drop the condition $e\neq 0$ in
the first integration and combine typical and atypical terms into
the single integral appearing in the first line. We observe that all
$\langle \cdot\rangle_{(z_0,y_0)}$ vanish on functions $e_0(e,n)$ with
$e=0$.
\smallskip

Let us now turn to the non-generic branes. These are localized
also in the fermionic directions. Hence, their density takes the
form
\begin{equation}
\rho^s_{z_0} \ = \ \, (-1)^s\,\delta(y-2\pi s)\,
 \delta(x-z_0)\, \delta (\etap)\, \delta(\etam)
\end{equation}
where $s$ is an integer. When this density is inserted into the
general prescription \eqref{eq:coinv}, we obtain another family of
co-invariants. Its expansion in terms of Ishibashi states reads
\begin{equation} \label{eq:MSSbst2}
\begin{split}
\langle\  \cdot\  \rangle^s_{z_0}& = \ _0\langle z_0;s| \ = \
  \int dedn
\,\frac{1}{\sqrt{e}}\, e^{2\pi i(n-1/2)s + iez_0} \ _0\llangle e,n|
 \\[2mm]
& \hspace*{-2cm} = \ \int_{e \neq 0} de dn \,\frac{1}{\sqrt{e}}\,
e^{2\pi i(n-1/2) s + iez_0} \  _0\llangle e,n| - \int_0^1 dn\,
e^{2\pi i (n-1/2) s} \  _0\llangle n|\ \ .
\end{split}
\end{equation}
Once more, the second line displays typical and atypical
contributions to the boundary state separately. In passing from
the first to the second line, we exploited $s \in \mathbb{Z}$
along with our observation \eqref{eq:nlimit}.
\smallskip

The two families $\langle \cdot \rangle_{(z_0,y_0)}$ with $y_0
\neq 2\pi s$ and $\langle \cdot \rangle^s_{z_0}$ are not entirely
independent. In fact, we note that boundary states from the
generic family may be `re-expanded' in terms of members from
the non-generic family when the paremeter $y_0$ tends to $2\pi
s$. The precise relation is
\begin{equation} \label{eq:branerel}
\lim_{y_0 \rightarrow 2\pi s} \, \langle f \rangle_{(z_0,y_0)} \ =
\ \frac{1}{i} \frac{\partial}{\partial z_0}
   \langle f \rangle^s_{z_0}\ \
\end{equation}
for all elements $f \in \L2$. We shall find that both
families of co-invariants can be lifted to the full field theory.
An analogue of relation \eqref{eq:branerel} also holds in the
field theory. It tells us that, for special values of the
parameters, branes from the generic family decompose into a
superposition of two branes from the non-generic family. Their
distance is finite for finite level but tends to zero as $k$ is
sent to infinity.

\section{Untwisted Boundary States and Their Spectra}

We are now prepared to spell out the boundary states and boundary
spectra for maximally symmetric branes with trivial gluing
conditions. As we have argued in the previous section, they come
in two different families. After a few comments on the relevant
Ishibashi states, we construct the boundary states for branes in
generic positions in the second subsection. Branes in non-generic
position are constructed in the third part of this section.

\subsection{Characters and Ishibashi states}

In this subsection we shall provide a list of untwisted Isibashi
states from which the boundary states of the \GL\ WZNW model will
be built in consecutive subsections. By definition, an untwisted
Ishibashi state is a solution of the following set of linear
relations
\begin{equation}
\left( X_n +\bar X_{-n} \right) |\Psi\rrangle\ = \ 0 \ \ \ \
\text{for} \ \ \ X = E,N,\Psi^\pm \ \ .
\end{equation}
These relation lift our invariance conditions \eqref{eq:MSSinv} from
the particle model to the full field theory. It is obvious that
solutions must be in one-to-one correspondence to invariants in
the minisuperspace theory.
\smallskip

To begin with, there exists a 2-parameter family of typical
Ishibashi states $|e,n\rrangle$ with $e \neq mk$ and $n \in
\mathbb{R}$. They can be uniquely characterized by their relative
overlaps
\begin{equation}
\llangle e,n| (-1)^{F^c} q^{L^c_0 - \frac{c}{24}} u^{N^c_0}
|e',n'\rrangle  \ = \ \delta(n'-n) \delta(e'-e) \ \chi_{\langle
e,n\rangle} (u,q)\ \
\end{equation}
where $L_0^c = (L_0 + \bar L_0)/2, N_0^c = (N_0 - \bar N_0)/2$ and
$\hat \chi_{\langle e,n\rangle}$ denotes the unspecialized
super-characters for typical representations. It takes the form
\begin{equation*}
    \hat{\chi}_{\langle e,n\rangle }(u,q)\ =\
    {u}^{n-1}{q}^{\frac{e}{2k}(2n-1+e/k)+1/8}
    \,\theta\Bigl(\mu-\frac{1}{2}({\tau}+1),{\tau}\Bigr)/\eta(\tau)^3
\end{equation*}
where $\mu$ is related to $u$ by $u = \exp(2\pi i \mu)$ and
similarly for $ q = \exp(2\pi i \tau)$, as usual. In comparison to
the minisuperspace theory we have set $u_1=1$ and $u_2 = u$. Since
$E_0$ and $\bar E_0$ are central the dependence on $u_1$ can be
re-introduced simply by multiplying the character functions with
$u_1^e$. When $e$ is a multiple of the level, $\hat{\chi}_{\langle
e,n\rangle}$ are the characters of reducible representations which
contain two atypical irreducible building blocks. As in the
particle theory, we shall also define $|mk,n\rrangle$ and
$\llangle mk,n|$ by a limiting procedure,
\begin{equation}\label{eq:atyplim1}
|mk,n\rrangle \ = \ \lim_{e \rightarrow mk}  \sin^{1/2}
 (\pi e/k) |e,n\rrangle\ \ \ , \ \ \
\llangle mk,n| \ = \ \lim_{e \rightarrow mk}  \sin^{1/2}
 (\pi e/k) \llangle e,n|\ \ .
\end{equation}
The Ishibashi states $|0,n\rrangle$ possess vanishing overlap
among each other and with the typical Ishibashi states.
\smallskip

In addition, we introduce a family of atypical Ishibashi states
$|n\rrangle^{(m)}$ and $^{(m)} \llangle n | $ for $n \in [0,1[, 
m \in \mathbb{Z}$. These correspond to the states $|n\rrangle_0$ 
and $_0\llangle n|$ that appeared in our discussion of the particle 
limit. Once more, we may characterize the Ishibashi states by their 
overlaps
\begin{equation}
^{(m)}\llangle n| (-1)^{F^c} q^{L^c_0 - \frac{c}{24}}
 u^{N^c_0}
|n'\rrangle^{(m)}  \ = \ \delta(n'-n) \delta(m-m')\
\hat\chi^{(m)}_{\langle n\rangle} (u,q)\ \ .
\end{equation}
Here, $\hat\chi^{(m)}_{\langle n\rangle}$ denotes the
unspecialized super-character of the atypical representation
$\langle n \rangle^{(m)}$, see Appendix A.3 for details, i.e.\ %
\begin{equation}
  \hat{\chi}_{\langle n\rangle}^{(m)}(u,q)
  \ =\ \frac{u^{n}}{1-zq^m}\,\frac{q^{\frac{m}{2}(m+2n+1)+1/8}
         \theta\Bigl(\mu-\frac{1}{2}(\tau+1),\tau\Bigr)}{\eta(\tau)^3}\ \ .
\end{equation}
It is important to stress that most atypical states are
obtained in eqs.\ \eqref{eq:atyplim1} as limits of typical
Ishibashi states.
\smallskip

To summarize, we have constructed a family of Ishibashi states
$|e,n\rrangle, e,n \in \mathbb{R}$, one for each Kac module of the
affine current algebra \agl. In addition, there is one `small'
family of Ishibashi states $|n\rrangle^{(m)}$ with $m \in \mathbb{Z}$
and $n \in [0,1[$. This second set of states is in one-to-one
correspondence with the set of atypical blocks of \agl.\footnote{Two atypical
irreducibles $\pi$ and $\pi'$ are said to be part of the same
block if there exists a sequence of irreps $\pi_0 = \pi, \pi_1,
\dots, \pi_{N-1}, \pi_N = \pi'$ such that any pair $\pi_i,
\pi_{i+1}$ of consecutive irreps in the sequence appears in the
composition series of some indecomposable. The two \agl\ %
representations $\langle n\rangle^{(m)} $ and $\langle
n\rangle^{(m)}$ are part of the same block whenever $m=m'$ and
$n-n' \in \mathbb{Z}$.}

\subsection{The generic boundary state}

In this section, we propose the boundary state corresponding to a
generic brane localized at $(z_0,y_0)$ with $y_0 \neq 2\pi s$ and perform a
non-trivial Cardy consistency check \cite{Cardy:1989ir}. Therefore,
we need to know the modular properties of the characters. They are
easily computed with the help of \cite{Mumford} and we list them in
appendix \ref{sc:Mods}.

\begin{prp}{\rm (Generic boundary state)} The boundary state of branes
associated with generic position parameters $z_0$, $y_0$ is
\begin{equation}\label{eq:bst1}
    |z_0,y_0\rangle\ =\ \sqrt{\frac{2i}{k}}\int de dn\
     \exp\bigl(i(n-1/2)y_0+iez_0\bigr)\ \sin^{1/2}(\pi e/k)\ |e,n\rrangle \ .
\end{equation}
We shall argue below that these boundary states give rise to
elementary branes if and only if the parameter
  $y_0 \not\in 2\pi\mathbb{Z}$.
\end{prp}
Before we show that our Ansatz for the generic boundary states
produces the expected boundary spectrum, let us make a few
comments. To begin with, it is instructive to compare the
coefficients of the Ishibashi states in $|z_0,y_0\rangle$ with the
minisuperspace result eq.\ \eqref{eq:MSSbst1}. If we send $k$ to
infinity, the factor $\sin^{1/2}(\pi e/k)$ is proportional to the
factor $\sqrt e$ that appears in the 1-point coupling of bulk
modes in the minisuperspace theory. The replacement $\sqrt e
\rightarrow \sin^{1/2} (\pi e/k)$ is necessary to ensure that the
field theory couplings are invariant under spectral flow. Let us
also stress that the integration in formula \eqref{eq:bst1}
extends over all $e$, including $e = mk$. Using our Ishibashi
states $|mk,n\rrangle$ from eq.\ \eqref{eq:atyplim1}, we may
rewrite the generic boundary states as
\begin{eqnarray*}
    |z_0,y_0\rangle& =& \sqrt{\frac{2i}{k}}\int_{e \neq mk} de dn\
     \exp\bigl(i(n-1/2)y_0+iez_0\bigr)\ \sin^{1/2}(\pi e/k)\
     |e,n\rrangle\\[2mm]
&  & \hspace{1.5cm} +\  \sqrt{\frac{2i}{k}} \sum_m \int dn
     \exp\bigl(i(n-1/2)y_0+imkz_0\bigr)\ |mk,n\rrangle\ \ .
\end{eqnarray*}
The second line displays explicitly how closed string states in
atypical representations couple to generic branes.
\smallskip

In order to check the consistency of our proposal for the boundary
states with world-sheet duality, we compute the spectrum between a
pair of generic branes,
\begin{eqnarray}
\nonumber \langle
z_0,y_0|(-1)^{F^c}\tilde{q}^{L_0^c}\tilde{z}^{N_0^c}|z'_0,y'_0\rangle
\!\!&\ =\ \!\!\frac{2i}{k}\int de' dn'
e^{i(n'-\frac12)(y'_0-y_0)+ie'(z'_0-z_0)}\sin(\pi e'/k)
   \hat{\chi}_{\langle e',n'\rangle }(\tilde{\mu},\tilde{\tau})\\[2mm]
&\ =\ \hat{\chi}_{\langle e,n\rangle }(\mu,\tau)\ -
                \hat{\chi}_{\langle e,n+1\rangle }(\mu,\tau)
\label{eq:gg}
\end{eqnarray}
where the momenta $e,n$ are related to the coordinates of the
branes according to
$$ e\ =\ \frac{k(y'_0-y_0)}{2\pi} \ \ \ \ , \ \ \
n\ =\ \frac{k(z'_0-z_0)}{2\pi}-\frac{y'_0-y_0}{2\pi}\ .$$
To begin
with, the result is a combination of characters with integer
coefficients. Hence, it can be consistently interpreted as the
partition function for open strings that stretch in between the
two branes. If we put both branes into the same position
$(z_0,y_0)$, then the result specializes to
\begin{equation} \label{eq:gbspec}\langle
z_0,y_0|(-1)^{F^c}\tilde{q}^{L_0^c}\tilde{u}^{N_0^c}
|z_0,y_0\rangle \ = \ \hat{\chi}_{\langle 0,0\rangle }(\mu,\tau)\
         -  \hat{\chi}_{\langle 0,1\rangle }(\mu,\tau)
                \ = \ \hat{\chi}_{{\mathcal P}_0}(\mu,\tau).
\end{equation}
In the last step we have observed that the super-characters of the
representation spaces over the two atypical Kac modules $\langle
0,0\rangle$ and $\langle 0,1\rangle'$ combine into the character of
the representation that is generated from the projective cover
${\mathcal{P}}_0$. This outcome was expected: it signals that the
state space of open strings on a generic branes contains no
bosonic zero modes and two fermionic ones. The latter give rise to
the four ground states of the projective cover. This is in
agreement with the fact that generic branes stretch out along the
fermionic directions.
\smallskip

There is one important subtlety in our interpretation of the
result \eqref{eq:gbspec} that we do not want to gloss over. While
the character of the projective cover $\hat{\mathcal{P}}_0$ is
the same as that of the two affine Kac modules, the corresponding
representations are not. The characters are blind against the
nilpotent parts in $L_0$ and hence they cannot distinguish between
an indecomposable and its composition series. But for the
conformal field theory, the difference is important. In
particular, the generator $L_0$ is diagonalizable on all Kac
modules, atypical or not, but it has a nilpotent contribution in
the \agl-module over $\mathcal{P}_0$. Hence, if the boundary
spectrum does transform in $\hat{\mathcal{P}}_0$, then some
boundary correlators are guaranteed to display logarithmic
singularities when two boundary coordinates come close to each
other. The information we obtained from the boundary states using
world-sheet duality alone is not sufficient to make any rigorous
statements on the existence of such logarithms. But in the
minisuperspace limit $k \rightarrow 0$ we have clearly identified
the projective cover $\mathcal{P}_0$ as the relevant structure.
Since $L_0$ is not diagonalizable in that limit, it cannot be so
for finite level $k$.

\subsection{Non generic point-like branes}

Let us now turn to the boundary states of non-generic untwisted
branes in the \GL\ WZNW model. From our discussion of the geometry
we expect them to be parametrized by a single real modulus $z_0$
and to possess a spectrum without any degeneracy of ground states.
These expectations will be met. Let us begin by spelling out the
formula for the non-generic boundary states.

\begin{prp}{\rm (Non-generic boundary states)}
The boundary states of elementary bra\-nes associated with
non-generic position parameters $z_0$ and $y_0=2\pi s, s \in
\mathbb{Z},$ are given by
\begin{equation}
    \begin{split}
        |z_0;s\rangle\ =\ \frac{1}{\sqrt{2ki}}
        \int dedn\ \exp\bigl(2\pi i (n-1/2) s + iez_0\bigr)\  \sin^{-1/2}(\pi e/k)\
         |e,n\rrangle\ .\\
    \end{split}
\end{equation}
\end{prp}
If we send the level $k$ to infinity in the boundary states
$|z_0;s\rangle$, then the coefficient of the Ishibashi
state $|e,s\rrangle$ gets replaced by $1/\sqrt{e}$ and thereby it
reproduces the coupling \eqref{eq:MSSbst2} of bulk modes in the
minisuperspace theory. Once more, the replacement
$1/\sqrt{e}\mapsto\sin^{-1/2}(\pi e/k)$ is necessary to ensure
spectral flow symmetry of the field theoretic couplings.%
\smallskip

Just like their cousins $|z_0;s\rangle_0$ in minisuperspace
(see eq.\ \eqref{eq:MSSbst2}), the boundary states $|z_0;s\rangle$ 
couple to atypical Ishibashi states, though this is again somewhat 
hidden in our notations. We can make this coupling more explicit 
by rewriting $|z_0;s\rangle$ in the form,  
\begin{equation} 
    \begin{split}
        |z_0;s\rangle & =\ \frac{1}{\sqrt{2ki}}
        \int_{e \neq mk} dedn\ \exp\bigl(2\pi i (n-1/2) s + 
   iez_0\bigr)\  \sin^{-1/2}(\pi e/k)\
         |e,n\rrangle\  \\[2mm] & \hspace*{2cm} - \, 
       \frac{1}{\sqrt{2ki}}
       \, \sum_{m} \int_0^1 dn\ \exp\bigl(2\pi i (n-1/2) s + 
   imk z_0\bigr)\ |n\rrangle^{(m)}\ \ . 
    \end{split}
\end{equation} 
Note that the non-generic boundary states only involve to the 
special family $|n\rrangle^{(m)}$ of atypical Ishibashi states. 
In case of generic boundary states, we had found non-vanishing 
couplings to the regular atypical Ishibashi states $|mk,n\rrangle$. 
\smallskip 

Let us verify that the proposed boundary states produce a
consistent open string spectrum. In order to do so, we investigate
the overlap between two non-generic boundary states
$|z_0;s\rangle$ and $|z_0';s'\rangle$,
\begin{eqnarray} \nonumber
\langle
z_0;s|(-1)^{F^c}\tilde{q}^{L_0^c}\tilde{z}^{N_0^c}|z'_0;s'\rangle
& = & \int  \frac{de'dn'}{2ki}\, \frac{e^{2\pi i(n'-1/2)(s'-s) +
ie'(z_0'-z_0)}}{\sin(\pi e'/k)} \ \hat{\chi}_{\langle
e',n'\rangle} (\tilde{\mu},\tilde{\tau})\\[2mm]
& = & \, \hat{\chi}_{\langle n \rangle}^{(m)}(\mu,\tau)
\label{eq:ngng}
\end{eqnarray}
where the labels $n$ and $m$ in the
character are related to the branes' parameters through
\begin{equation}
n \ = \ \frac{k(z'_0-z_0)}{2\pi} + s-s' \ \ \ \ , \ \ \ m = s'-s\ \ .
\end{equation}
$\hat{\chi}_{\langle n \rangle}^{(m)}$ are characters of
atypical irreducible representation of \agl. For $m=0$ the
corresponding representations are generated from the 1-dimensional
irreducible atypical representations $\langle n\rangle$ of the
finite-dimensional Lie superalgebra \gl\ by application of current
algebra modes. The representations with $m \neq 0$ are obtained
{}from those with $m=0$ by spectral flow (see Appendix A).
\smallskip

We also want to look at the spectrum of boundary operators that
can be inserted on a boundary if we impose non-generic boundary
conditions with parameters $z_0$ and $s$. Specializing eq.\
\eqref{eq:ngng} to the case with $z'_0 = z_0$ and $s'=s$ we find
\begin{equation} \nonumber
\langle
z_0;s|(-1)^{F^c}\tilde{q}^{L_0^c}\tilde{u}^{N_0^c}|z_0;s\rangle
  \ =\ \hat{\chi}_{\langle 0 \rangle}^{(0)}(\mu,\tau) \ \ .
\end{equation}
Hence, the spectrum consists of states that are generated from a
single invariant ground state $|0\rangle$ by application of
current algebra modes with negative mode indices. In particular,
the zero modes of the fermions act trivially on ground states.
This is in agreement with our geometric insights according to
which non-generic branes are localized in all directions,
including the two fermionic ones.
\smallskip

We may now ask what happens if we send the parameter $y_0$ of the
generic brane to $y_0=2\pi s$. From our formulas for boundary
states we deduce that
\begin{equation*}
|z_0,2\pi s\rangle\ =\ %
\int \frac{dedn}{\sqrt{2ki}}\, \frac{e^{ie(z_0+\frac{\pi}{k})}
     - e^{ie(z_0-\frac{\pi}{k})}}{\sin^{1/2}(\pi e/k)} \,
e^{2\pi i(n-1/2)s}\, |e,n\rrangle
\ =\ |z_0+\pi/k;s\rangle -
|z_0  - \pi/k;s\rangle\ .
\end{equation*}
In other words, when a generic brane is moved onto one of the
special lines $y_0 = 2\pi s$, it decomposes into a
brane-anti-brane pair. Its constituents sit in positions $z_0\pm
\pi/k$ and possess the same discrete parameter $s$. This relation
between non-generic branes and generic branes in non-generic
positions is a field theoretic analogue of the equation
\eqref{eq:branerel} we discovered in the minisuperspace
theory.%

\section{Comparison with Cardy's Theory}

Let us recall a few rather basis facts concerning branes in
rational unitary conformal field theory. For simplicity we shall
restrict to cases with a charge conjugate modular invariant and a
trivial gluing automorphism $\Omega$ (the so-called `Cardy case').
This will allow a comparison with the results of the previous
subsections. In the Cardy case, elementary boundary conditions
turn out to be in one-to-one correspondence with the irreducible
representations of the chiral algebra \cite{Cardy:1989ir}.
Let us label these by $J$,
with $J=0$ being reserved for the vacuum representation. The
boundary condition with label $J=0$ has a rather simple spectrum
containing only the vacuum representation $\mathcal{H}_0$. More
generally, if we impose the boundary condition $J=0$ on one side of
the strip and any other elementary boundary condition on the
other, the spectrum consists of a single irreducible ${\cal H}_J$.
Finally, the spectrum between two boundary conditions with label
$J_1$ and $J_2$ is determined by the fusion of $J_1$ and $J_2$. We
shall now discuss that all these statements carry over to
untwisted branes in the \GL\ WZNW model. The fusion procedure,
however, can provide spectra containing indecomposables that are
not irreducible.
\smallskip

\subsection{Brane parameters and representations}

We proposed that the \GL\ WZNW model possesses two families of
elementary branes. The first one is referred to as the generic
family and its members are parametrized by $(z_0,y_0)$ with $y_0
\neq 2\pi s, s \in \mathbb{Z}$. Boundary states for the generic
branes were provided in subsection 4.2. These are also defined for
integer $y_0/2\pi$ but we have argued that the corresponding
branes are not elementary. They rather correspond to
superpositions of branes from the second family. This second
family consists of branes with only one continuous modulus $z_0$
and a discrete parameter $s$. Their boundary states can be found
in subsection 4.3.
\smallskip

There is one distinguished brane in this second family with
$z_0=0$ and $s=0$. We propose that it plays the role of the $J=0$
brane in rational conformal field theory. In order to confirm this
idea, we compute the spectrum of open strings stretching between
$z_0=0,s=0$ and any of the other elementary branes. If the second
brane is non-generic with parameters $z_0,s$, the relative
spectrum reads
\begin{equation}
\langle 0;
0|(-1)^{F^c}\tilde{q}^{L_0^c}\tilde{u}^{N_0^c}|z_0;s\rangle\ =\
\hat{\chi}_{\langle n\rangle}^{(m)}(\mu,\tau)\
\end{equation}
where the parameter $n$ on the character is
\begin{equation} \label{eq:posrep1}
 n \ = \ n(z_0;s) \ =\ \frac{kz_0}{2\pi} - s \ \ \ , 
\ \ \ m \ =\ m(z_0;s) \ = \ s\ .
\end{equation}
Indeed, we see that the open string spectrum corresponds to a
single irreducible atypical module of \agl, in agreement with the
expectations from rational conformal field theory.%
\smallskip

Let us now consider the case in which the second brane is located
in a generic position $(z_0,y_0)$. From the boundary state we find
\begin{equation}
\langle0;
0|(-1)^{F^c}\tilde{q}^{L_0^c}\tilde{u}^{N_0^c}|z_0,y_0\rangle \ =
\ \hat{\chi}_{\langle e,n\rangle }(\mu,\tau)\ ,
\end{equation}
where the parameters of the character on the right hand side are
\begin{equation} \label{eq:posrep2}
 e \ = \ e(z_0,y_0) \ = \ \frac{ky_0}{2\pi}  \ \ \ , \ \ \
 n \ = \ n(z_0,y_0) \ =\ \frac{k z_0}{2\pi}-\frac{y_0}{2\pi}+ \frac12\ \ .
\end{equation}
As long as $y_0/2\pi$ is not an integer, $e$ is not a multiple of
the level and therefore, $\hat \chi_{\langle e,n\rangle}$ is the
character of a single irreducible representation of \agl.
\smallskip

At this point we have found that all our elementary branes are
labelled by irreducible representations of \agl. In case of the
elementary generic branes, the relation between the position
moduli $(z_0,y_0), y_0 \neq 2 \pi m,$ and representation labels
$\langle e,n\rangle, e \neq mk,$ is provided by eq.\
\eqref{eq:posrep2}. All typical irreducible representations of
\agl\ appear in this correspondence. For the non-generic branes the
relation between their parameters $(z_0;s)$ and the representation
labels of an atypical irreducible can be found in eq.\
\eqref{eq:posrep1}. Once more, all atypical irreducibles appear in
this correspondence. Hence, branes in the \GL\ WZNW model are in
one-to-one correspondence with irreducible representations of the
current superalgebra \agl, just as in rational conformal field
theory.

\subsection{Brane spectra and fusion}

Let us now analyze whether we can find the spectrum between a pair
of elementary branes through fusion of the corresponding current
algebra representations. For the convenience of the reader we have
listed the relevant fusion rules for irreducible representations
of the current superalgebra \agl\ in Appendix~\ref{sc:AffFus}.
\smallskip

The spectrum between two typical branes with parameters
$(z_0,y_0)$ and $(z'_0,y'_0)$ has been computed in eq.\
\eqref{eq:gg}. We can convert the brane parameters into
representation labels with the help of eq.\ \eqref{eq:posrep2} and
then exploit the known fusion product of the corresponding
representations. In case $y'_0-y_0 \neq 2\pi \mathbb{Z}$ we
find
\begin{eqnarray} \label{fus1} & &  \Bigl\langle
\frac{ky_0}{2\pi},\frac{kz_0}{2\pi}-\frac{y_0}{2\pi} + \frac12
\Bigr\rangle^\ast \otimes_F \Bigl\langle
\frac{ky'_0}{2\pi},\frac{kz'_0}{2\pi}-\frac{y'_0}{2\pi} +
\frac12
\Bigr\rangle \\[4mm] & & \hspace*{.5cm} \cong \ \Bigl\langle \frac{k(y'_0-y_0)}{2\pi},
\frac{k(z'_0-z_0)}{2\pi} - \frac{y'_0-y_0}{2\pi} + 1\Bigr\rangle \,
\oplus \, \Bigl\langle \frac{k(y'_0-y_0)}{2\pi},
\frac{k(z'_0-z_0)}{2\pi} - \frac{y'_0-y_0}{2\pi}\Bigr\rangle' \nonumber
\end{eqnarray}
Here, $\otimes_F$ denotes the fusion product and we used the rule
$\langle e,n\rangle^\ast=\langle -e,-n+1\rangle'$ for the
conjugation of representations. Then we inserted the known fusion
rules while keeping track of whether the representation is
fermionic or bosonic. The result agrees nicely with the true
spectrum we computed earlier.
\smallskip

When the difference $(y'_0-y_0)/2\pi = m$ is an integer, the
fusion of the two representations on the left hand side of
\eqref{fus1} results in a single indecomposable. It is the image
of the affine representation over the projective cover
$\hat{\mathcal{P}}_{(k(z'_0-z_0)-(y'_0-y_0))/2\pi}$ under $m$ units of
spectral flow, i.e.\ %
\begin{equation} \label{fus2} \Bigl\langle
\frac{ky_0}{2\pi},\frac{kz_0}{2\pi}-\frac{y_0}{2\pi} + \frac12
\Bigr\rangle^\ast \otimes_F \Bigl\langle
\frac{ky'_0}{2\pi},\frac{kz'_0}{2\pi}-\frac{y'_0}{2\pi} +
\frac12 \Bigr\rangle \ = \ \Bigl(\mathcal{P}^{(m)}_{(k(z'_0-z_0)-(y'_0-y_0))/2\pi}\Bigr)'
\end{equation}
where $m = (y'_0-y_0)/2\pi$. Our minisuperspace theory along with
the boundary states confirm this result in the case $y_0 = y_0'$
and $z_0 = z_0'$ (see our discussion at the end of section 4.2).
For other choices of the parameters, we only see that the fusion
rules provide a representation with the correct character. Whether
the true state space is given by a single indecomposable or by a
sum of Kac modules or even irreducibles cannot be resolved
rigorously with the methods we have at our disposal. Nevertheless,
it seems very likely that the projective cover is what appears
since this is the only result which is also consistent with
spectral flow symmetry.
\smallskip

The fusion between atypical irreducibles is rather simple. It
leads to a prediction  for the spectrum between two non-generic
branes that should be checked against our earlier result
\eqref{eq:ngng},
$$ \Bigl(\,\Bigl\langle \frac{kz_0}{2\pi} -s\Bigr\rangle^{(s)}\Bigr)^\ast \otimes_F
   \,\Bigl\langle \frac{kz'_0}{2\pi}-s'\Bigr\rangle^{(s')} \ \cong \
   \,\Bigl\langle \frac{k(z'_0-z_0)}{2\pi}+s-s'\Bigr\rangle^{(s'-s)}\ \ .
$$
Once more, the findings from world-sheet duality are consistent
with the fusion prescription. There is one final check to be
performed. It concerns the spectrum between a non-generic brane
with parameters $(z_0;s)$ and a generic one with moduli
$(z_0,y_0)$. From the fusion we find
\begin{equation}
\Bigl(\Bigl\langle \frac{kz_0}{2\pi} -s \Bigr\rangle^{(s)}\Bigr)^\ast
  \otimes_F\Bigl\langle
\frac{ky'_0}{2\pi},\frac{kz'_0}{2\pi}-\frac{y'_0}{2\pi} + \frac12
\Bigr\rangle \ = \ \Bigl\langle -s k + \frac{ky'_0}{2\pi},
\frac{k(z'_0-z_0)}{2\pi} - \frac{y_0'}{2\pi}+s+\frac12\Bigr\rangle
\ \ .
\end{equation}
It may not come as a big surprise that this
fusion rule correctly predicts the spectrum between a generic and
a non-generic brane. In fact, from our formulas for boundary
states and modular transformation we find
\begin{equation}
  \begin{split}
& \bigl\langle z_0;s\bigr| (-1)^{F^c}\tilde q^{L_0^c} \tilde
u^{N^c_0} \bigl|z'_0,y_0'\bigr\rangle \ = \ \hat \chi_{\langle e
,n \rangle}(\mu,\tau)
\\[4mm] \text{where}  \ &\ \ \ e \ = \ - k s +
\frac{ky_0'}{2\pi}\ \ , \ \ n \ =  \frac{k(z'_0-z_0)}{2\pi} -
\frac{y_0'}{2\pi}+s + \frac12\ .
  \end{split}
\end{equation}
In conclusion we found that the spectra between any pair of
elementary branes may be determined by the fusion of the
corresponding irreducible representations. It is important to
stress that the fusion product of irreducible representations can
produce representations that are not fully reducible.

\section{Twisted Brane: Geometry and Boundary State}

This final section contains a brief discussion of twisted branes.
By definition, twisted branes in the \gl\ model preserve one copy
of the affine Lie superalgebra \agl. The construction of the
relevant generators differs from the case of untwisted branes by
the action of an outer (gluing) automorphism $\Omega$ on
anti-holomorphic bulk currents. We shall find that there is a
single twisted brane boundary condition corresponding to a brane
which extends in both bosonic and fermionic directions. As for
untwisted branes, we shall first extract the brane's geometry from
the gluing conditions. Thereafter, we study the unique Ishibashi
and boundary state in the particle limit. Finally, the
minisuperspace results are lifted to the full field theory.
\smallskip

In the case of the automorphism $\Om$, we can easily bring the
associated gluing conditions \eqref{eq:glue} for super-currents
into the form
\begin{equation}
    \begin{split}
        \n Y =\ 0 \ \hspace*{1.5cm} , \ \ \ \
        \n\bar{\xi} &=\ ie^{-iY}\p\xi\\[2mm]
	\n X  - \ 2ie^{iY}\xi\n\bar{\xi} =\ 0\ \ \ \ \ , \label{twglue}
        \ \ \ \ \n\xi & = \  -ie^{iY}\p\bar{\xi} \ \ ,
    \end{split}
\end{equation}
for all $z=\bar{z}$. Here, we have redefined the fermionic fields
$\xi=\frac{e^{iY}}{2}(\cp+\cm)$ and
$\bar{\xi}=\frac{1}{2} (\cm-\cp)$. The bosonic fields, on
the other hand, remain unaltered. This parametrization is motivated by a new choice of coordinates on the supergroup \GL
\begin{equation}
	\begin{split}
                g\ =\ e^{i\cm\Pm}e^{ixE+iyN}e^{i\cp\Pp}\ &=\
		e^{i\bar{\xi}\Pm}e^{-i\xi\Pp}e^{izE+iyN}e^{-i\xi\Pm}e^{-i\bar{\xi}\Pp}\\
		&=\ \Omega(e^{i\bar{\xi}\Pp}e^{i\xi\Pm})e^{izE+iyN}e^{-i\xi\Pm}e^{-i\bar{\xi}\Pp}\ \\
	\end{split}
\end{equation}
which is obtained by twisted conjugation of bosonic elements with fermionic ones.
\smallskip

We can re-express the metric and H-field in terms of the new
coordinates $x,y,\xi\,\bar \xi$,
\begin{equation*}
\begin{split}
	ds^2 & = \ 2dxdy+4d\xi d\bar{\xi}-4i\xi dyd\bar{\xi},\\[2mm]
 H& = \ 2ie^{-iy}d\xi\wedge d\xi\wedge dy
 -2ie^{iy}d\bar{\xi}\wedge d\bar{\xi}\wedge dy)\ \ .
\end{split}
\end{equation*}
Using our expression for the metric we infer the following formula
for the B-field from our gluing conditions \eqref{twglue},
\begin{equation*}
B \ =\ -2e^{-iy}d\xi\wedge d\xi - 2e^{iy}d\bar{\xi}\wedge
d\bar{\xi}\ \ .
\end{equation*}
It is straightforward to verify that  that $dB=H$. We conclude
that twisted branes are stretched out into all directions of our
supergroup.

Consequently, the space of functions on a twisted D-brane is given
by $\L2$. Since twisted branes admit an action of \GL\ the space
of functions carries an action of the Lie superalgebra \gl, namely the
twisted adjoint action $\ad^{\Omega}_X = R_X + L^\Omega_X$ where
\begin{equation*}
L_E^\Omega\ =\ i\del_x \ , \ \ L_N^\Omega\ =\ i\del_y+\etap\del_+ \ , \ \
L_-^\Omega\ =\ -\del_+ \ , \ \ L_+^\Omega\ =\
e^{-iy}\del_--i\etap\del_x \ \ .
\end{equation*}
The generators $R_X$ are given by the same formulas as above.
Analyzing the representation content of $\L2$ we then find three
different kinds of representations. These include the typicals
$\langle -2k,-2l+1\rangle$ which are generated by
$e_0(k,l)=\exp(ikx+ily), \xi e_0(k,l-1)$. We recall that in our
conventions for $\langle e,n\rangle$ the state with smaller $N$
eigenvalue is taken to be bosonic. Furthermore, there exist
typicals $\langle -2k,-2l+2\rangle'$ generated by
$\bar{\xi}e_0(k,l),e_0(k,l-1)+2k\xi\bar{\xi}e_0(k,l-1)$. In this case, the
state with lower $N$ eigenvalue is fermionic, hence the prime $'$.
Finally, representations with vanishing eigenvalue of $E$
decompose into projective covers of atypicals. In summary, under
the twisted adjoint action, the space of functions decomposes as
\begin{equation*}
\L2^{\text{twisted}}\ \cong\  \int_{e\neq 0}dedn\, \Bigl[\, \langle
e,n\rangle \oplus \langle e,n\rangle' \,\Bigr]\, \oplus\, \int dn\,
\mathcal{P}_n\ \ .
\end{equation*}
We see that fermionic and bosonic states with any given
eigenvalue of $E$ and $N$ come in pairs. Therefore, the supertrace
of $u_1^{L^\Omega_E-R_E} u_2 ^{L^\Omega_N-R_N}$ vanishes
identically.
\smallskip

Concerning the construction of minisuperspace Ishibashi states
$|\psi\rrangle^\Omega_0$ satisfying the twisted invariance
condition
\begin{equation}
\left( R_X + L_X^\Omega \right) |\psi\rrangle^\Omega_0 \ = \ 0
\end{equation}
we observe that the space of functions on \GL\ contains a single
element invariant under the twisted adjoint action, namely the
constant function
$$|0\rrangle^\Omega_0 \ = \ e_0(0,0)\ \ \ . $$
Its dual is given by
$$ ^\Omega_0\llangle 0| \ = \ \int d\mu\,e_0(0,0) = \int d\mu\ \ .$$
The linear functional $^\Omega_0\llangle 0|$ is indeed the unique
twisted co-invariant on \GL. We note that $|0\rrangle^\Omega_0$
and $^\Omega_0\llangle 0|$ possess vanishing overlap, i.e.\
$$  ^\Omega_0\llangle 0|(-1)^F u_1^{L^\Omega_E - R_E} u_2^{L^\Omega_N - R_N}
 |0\rrangle^\Omega_0 \ = \ 0 $$
simply because the relevant integrand contains no fermionic zero
modes.
\smallskip

Having the semi-classical Ishibashi state at our disposal, we can
turn to the boundary state. Our geometric interpretation of
twisted branes suggests that their semi-classical density is given
by $\rho(x,y,\xi,\bar\xi)=1$, corresponding to a brane that fills
the entire target space. We see that
$$ \langle\ \cdot\  \rangle^\Omega \ = \ _0\langle \Omega |
\ = \ \int d\mu \ = \ ^\Omega_0\llangle 0 | \ \ .
$$
All this lifts straightforwardly to the full field theory. We
obtain unique Ishibashi states $|0\rrangle^\Omega$ and $^\Omega
\llangle0|$ which we can identify with the boundary states,
$$ | \Omega \rangle \ = \ |0\rrangle^\Omega \ \ \ , \ \ \ \
   \langle \Omega | \ = \ ^\Omega \llangle 0| $$
just as in the case of Neumann boundary conditions for a free
uncompactified boson. The interaction between two such branes is
encoded in the overlap
$$ \ \langle \Omega |(-1)^{F^c} \tilde{q}^{L^c_0} u^{N^c_0}
|\Omega\rangle \ = \ 0 \ \ , $$
where $N^c_0 = \bigl(\Omega(N_0)- \bar
N_0\bigr)/2$. Through the modular bootstrap, vanishing of this overlap
implies that the boundary partition function vanishes as well. In
our minisuperspace approximation we did observe already that
contributions from bosonic and fermionic states to the partition
function cancel each other. The same holds true for the full field
theory since creation operators also come in pairs. Hence, our
results are consistent with the world-sheet duality.

Admittedly, the simplest version of the  modular bootstrap does
not constrain the form of our boundary states very significantly.
But there exists more stringent tests, such as bootstrap relations
involving the overlap between twisted and untwisted D-branes
\cite{Fuchs:1999zi,Birke:1999ik}. We have no doubt that these can be
worked out to confirm our proposal for the twisted boundary state.

\section{Conclusions}

In this work we have studied maximally symmetric branes in the
WZNW model on the simplest supergroup \GL . Following previous
reasoning for bosonic models \cite{Alekseev:1998mc} we have shown
that such branes are localized along (twisted) super-conjugacy
classes, an insight that generalizes straightforwardly to other
supergroup target spaces. As in the case of the $p=2$ triplet 
theory \cite{Gaberdiel:2006pp}, untwisted branes turn out 
to be in one-to-one correspondence with irreducible representations 
of the current algebra. This correspondence relies on the existence 
of an `identity' brane whose spectrum consists of the irreducible 
vacuum representation only. The spectrum between the identity and any
other elementary brane is built from a single irreducible of \agl\
and any such irreducible appears in this way. Moreover, one can
compute the spectrum between any two elementary branes by fusion
of affine representations. What we have just listed are
characteristic features of Cardy's theory for rational
non-logarithmic conformal field theories. Our work proves that
they extend at least to one of the simplest logarithmic field
theory and it seems very likely that they hold more generally in
all WZNW models on (type I) supergroups, see also 
\cite{Gaberdiel:2006pp} for related findings in the $p=2$ 
triplet theory. 
\smallskip

In spite of these parallels to bosonic WZNW models, branes on
supergroups possess a much richer spectrum of possible geometries.
Whereas Dirichlet branes on a purely bosonic torus, for example,
are all related by translation, we discovered the existence of
atypical lines on the bosonic base of the \GL\ WZNW model. The
distance between any two such neighboring parallel lines is
controlled by the level $k$. When a typical untwisted brane is
moved onto one of these lines, it splits into two atypical ones.
Individual atypical branes possess a single modulus that describes
their dislocation along the atypical lines. In order for them to
leave an atypical line they must combine with a second atypical
brane. Processes of this kind model the formation of long
multiplets from shorts. Hence, on more general group manifolds,
more than just two atypical branes may be required to form a
generic brane. Let us stress, however, that the notions of long
(typical) and short (atypical) multiplets which are relevant for
such processes derive directly from the representation theory of the affine
Lie superalgebra. Thereby, all spectral flow symmetries are
built into our description. We also wish to point out the obvious
similarities with so-called fractional branes at orbifold singularities, 
see e.g.\ the discussions in section 4.3 of \cite{Recknagel:1998ih}.  
\smallskip

Another interesting and new feature of branes on \GL\ is the
occurrence of boundary spectra that cannot be decomposed into a
direct sum of irreducibles. In particular we have shown that the
spectrum of boundary operators on a single generic brane is
described by the projective cover of the vacuum module.
For more general group manifolds, we expect the corresponding
projective cover to be present as well, though along with
additional stuff. The generator $L_0$ of dilatations is not 
diagonalizable on projective covers, see e.g.\ 
\cite{Schomerus:2005bf}. According to the usual reasoning, this 
implies the existence of logarithmic singularities in boundary 
correlation functions on branes in generic positions. As we have 
remarked before, the modular bootstrap alone did not allow for such a
strong conclusion as it is blind to all nilpotent contributions
within $L_0$. But in addition to the standard conformal field
theory analysis, our investigation of the \GL\ WZNW model also
draws from the existence of the geometric regime at large level
$k$. The presence of projective covers is easily understood in the
minisuperspace theory and it must persist when field theoretic
corrections are taken into account. \smallskip

There are a few obvious extensions of the above analysis that seem to merit
closer investigation. These include the computation of  boundary
correlation functions for twisted and untwisted branes in the \GL\
model. We expect that correlators with a small number of bulk
and/or boundary insertions may be computed using free field
techniques, as in the case of bulk models
\cite{Schomerus:2005bf,Gotz:2006qp}. It would also be interesting
to study the various brane geometries that can come up on other
supergroup manifolds. We plan to report on both issues in the near
future.
\bigskip

\noindent {\em Note added:} While we were in the final stages of
preparing this manuscript, a related paper \cite{Gaberdiel:2007jv}
appeared which discusses branes in triplet models with $p \geq 2$. 
The results of Gaberdiel and Runkel show that branes in triplet 
models share many features with the outcome of our analysis. In 
particular, for trivial gluing automorphism, branes in both models 
are labelled by irreducible representations of the chiral algebra. 
Also the labels for relevant Ishibashi states follow the same 
pattern: We have found one `generic' Ishibashi state for each 
Kac module and an exceptional family with members being associated 
to atypical blocks. When the same rules are applied to the triplet 
models, we obtain a set of Ishibashi states that seems closely 
related to those used in \cite{Gaberdiel:2007jv}. Furthermore, 
Gaberdiel and Runkel also find that the partition function
for any pair of boundary conditions may be determined by 
fusion of representations. The existence of a geometric 
regime for the \GL\ WZNW model allows us to go one step 
further. It gives us full control over the structure of the 
state space and thereby also over the nilpotent contributions 
to $L_0$ which are not visible in partition functions. Fusion 
of \agl\ representations was shown to correctly reproduce the 
state spaces of boundary theories in the \GL\ WZNW model. Let 
us stress, however, that the triplet and the \GL\ WZNW model 
are close cousins (see e.g. the discussion in \cite{Quella:2007hr}). 
It would therefore be somewhat premature to claim that all these 
structures will be present in more general logarithmic 
conformal field theories.

\subsubsection*{Acknowledgements}

  We are grateful to Matthias Gaberdiel, Gerhard G\"otz, Ingo
  Runkel, Hubert Saleur and Aliosha Semikhatov for useful discussions and
  comments. This research was supported in part by the EU Research Training
  Network grants 
  ``ForcesUniverse'' (contract number MRTN-CT-2004-005104), ``Superstring
  Theory'' (contract number MRTN-CT-2004-512194) 
  and by the PPARC rolling grant
  PP/C507145/1. Part of this work has been performed while Thomas Quella
  was working at King's College London, funded by a PPARC postdoctoral
  fellowship under reference PPA/P/S/2002/00370.

\appendix
\section{\label{sc:App} The Representation Theory of $\mathbf{\widehat{gl}(1|1)}$}

\subsection{\label{sc:SpecFlow}Spectral flow automorphisms}

   A useful tool for the investigation of the current algebra
   $\widehat{gl}(1|1)$ and its representations are spectral flow
   automorphisms. The first one, $\gamma_m$, leaves the modes $N_n$
   invariant and acts on the remaining ones as
\begin{equation}
   \label{eq:SFE}
   \gamma_m(E_n)
   \ =\ E_n+km\delta_{n0} \ , \ \ \
\gamma_m(\Psi^\pm_n)\ =\ \Psi^\pm_{n\pm m}\ \ .
\end{equation}
   The previous transformation also induces a modification
   of the energy momentum tensor which is determined by
\begin{equation}
   \gamma_m(L_n)\ =\ L_n+mN_n\ \ .
\end{equation}
   Since the rank of \GL\ is two, there is a second one
   parameter family of spectral flow automorphisms
   $\tilde{\gamma}_\zeta$ which is parametrized by a continuous
   number $\zeta$. It is rather trivial in the sense that its
   action does not act on the mode numbers,
\begin{equation}
   \tilde{\gamma}_\zeta(N_n) \ =\ N_n+k\,\zeta\,\delta_{n0} \ \ \text{and}
   \ \ \tilde{\gamma}_\zeta(L_n) \ =\ L_n+\zeta\,E_n\ \ .
\end{equation}
   All other modes of the currents are left invariant.
\smallskip

   The two spectral flow symmetries above induce a map on
   the set of representations of \agl. Given any
   representation $\rho$ we obtain two new ones by defining
   $\rho_m=\rho\circ\gamma_m$ and
   $\tilde{\rho}_\zeta=\rho\circ\tilde{\gamma}_\zeta$. The latter
   is not very exciting but the former will play a crucial
   role below. Let us thus state in passing that the
   super-characters of these representations are related by
\begin{equation}
   \chi_{\rho_m}(\mu,\tau)\ =\ \chi_\rho(\mu+m\tau,\tau)\ \ .
\end{equation}
   This formula gives severe restrictions on the nature
   of the representations $\rho_m$.

\subsection{\label{sc:Theta}Some formulas concerning Theta functions}

   Let us recall some facts about the theta function in one variable,
   the reference is Mumford's first book \cite{Mumford}.
   $\theta(\mu,\tau)$ is the unique holomorphic function on
   $\mathbb{C}\times\mathbb{H}$, such that
\begin{equation}
   \begin{split}
     \theta(\mu+1,\tau)
     &\ =\ \theta(\mu,\tau),\\[2mm]
     \theta(\mu+\tau,\tau)
     &\ =\ e^{-\pi i\tau}e^{-2\pi i \mu}\theta(\mu,\tau),\\[2mm]
     \theta(\mu+\frac{1}{2},\tau+1)
     &\ =\ \theta(\mu,\tau),\\[2mm]
     \theta(\mu/\tau,-1/\tau)
     &\ =\ \sqrt{-i\tau}e^{\pi i\mu^2/\tau}\theta(\mu,\tau)\\[2mm]
     \lim_{\text{Im}(\tau)\rightarrow \infty}\theta(\mu,\tau)
     &\ =\ 1\ \ .
   \end{split}
\end{equation}
   The theta functions has a simple expansion as an infinite product,
\begin{equation}
   \theta(\mu,\tau)
   \ =\ \prod_{m=0}^\infty\bigl(1-q^m\bigr)
        \prod_{n=0}^\infty\bigl(1+u^{-1}q^{n+1/2}\bigr)\bigl(1+uq^{n+1/2}\bigr)\
\ ,
\end{equation}
   where $q=e^{2\pi i\tau}$ and $u=e^{2\pi i \mu}$. The \agl\ characters
   in the RR sector we shall present in the next section have a simple
   expression in terms of the variant
\begin{equation}
   \theta\Bigl(\mu-\frac{1}{2}(\tau+1),\tau\Bigr)
   \ =\
(1-u)\prod_{n=1}^\infty\bigl(1-q^n\bigr)\bigl(1-uq^n\bigr)\bigl(1-u^{-1}q^n\bigr)\
\ .
\end{equation}
   Its behavior under modular $S$ transformations which send the
   arguments of the theta function to $\tilde{\tau}=-1/\tau$ and
   $\tilde{\mu}=\mu/\tau$ can be deduced from the properties above. One
   simply finds
\begin{equation}
   \theta\Bigl(\tilde{\mu}-\frac{1}{2}(\tilde{\tau}+1),\tilde{\tau}\Bigr)
   \ =\ i\sqrt{-i\tilde{\tau}}\,e^{\pi i\tilde{\mu}^2/\tilde{\tau}}\,
        u^{1/2}\tilde{u}^{-1/2}\,q^{-1/8}\tilde{q}^{1/8}\,
        \theta\Bigl(\mu-\frac{1}{2}(\tau+1),\tau\Bigr)\ \ .
\end{equation}

\subsection{\label{sc:Reps}Representations and their characters}

   In this appendix we review the representations of the current
   superalgebra \agl\ that are relevant for our discussion in the main
   text. We shall slightly deviate from the presentation in
   \cite{Schomerus:2005bf} in putting even more emphasis on the role of
   the spectral flow automorphism \eqref{eq:SFE}. The latter is the
   only constituent which leads to a substantial difference between the
   representation theory of the finite dimensional subalgebra \gl\ and
   that of its affinization \agl.
\smallskip

   All irreducible representations of \agl\ are quotients of Kac
   modules. Just as for \gl, we distinguish between Kac modules
   $\langle e,n\rangle$ and anti Kac modules $\overline{\langle
     e,n\rangle}$. These symbols have been chosen since the ground
   states transform in the corresponding representations of the
   horizontal subalgebra \gl.\footnote{We would like to stress that the
   representations $\langle mk,n\rangle$ and $\overline{\langle
     mk,n\rangle}$ are inequivalent for $m\in\mathbb{Z}$ even though
   their ground states transform identically as long as $m\neq0$. The
   reason becomes clear below.} For $e\not\in k\mathbb{Z}$ both types
   of representations will be called typical, otherwise
   atypical. Typical representations are irreducible and one has the
   equivalence $\langle e,n\rangle\cong\overline{\langle
     e,n\rangle}$. The super-character of (anti) Kac modules can easily
   be found to be
\begin{equation}
   \label{eq:CharTyp}
   \hat{\chi}_{\langle e,n\rangle}(\mu,\tau)
   \ = \ \hat{\chi}_{\overline{\langle e,n\rangle}}(\mu,\tau)
   \ = \ u^{n-1}q^{\frac{e}{2k}(2n-1+e/k)+1/8}
   \theta\Bigl(\mu-\frac{1}{2}(\tau+1),\tau\Bigr)\bigr/\eta(\tau)^3\ \ .
\end{equation}
   When writing down this expression we assumed the ground state with
   quantum numbers $(E_0,N_0)=(e,n)$ to be fermionic. The spectral flow
   $\gamma_m$ transforms the characters of Kac modules according to
\begin{equation}
   \label{eq:SFKac}
   \gamma_m: \quad \chi_{\langle e,n\rangle }(\mu,\tau)
             \ \mapsto\ (-1)^m\chi_{\langle e+mk,n-m\rangle }(\mu,\tau)\ \ .
\end{equation}
   This equation should be interpreted as defining a map between
   representations. We recognize that $\langle e,n\rangle$ is
   transformed into $\langle e+mk,n-m\rangle$ under $\gamma_m$ and that
   the parity of the module is changed if $m$ is odd. A change of
   parity occurs if the interpretation of what are bosonic and what
   are fermionic states is altered compared to the standard choice.
\smallskip

   The equivalence between Kac modules and anti Kac modules is
   destroyed for $e\in k\mathbb{Z}$. For these values the
   representations $\langle mk,n\rangle$ and $\overline{\langle
     mk,n\rangle}$ degenerate and exhibit a single singular vector
   which can be found on energy level $|m|$, see
   \cite{Schomerus:2005bf} for details.\footnote{In order to avoid
   confusion we would like to emphasize that the construction in
   \cite{Schomerus:2005bf} gives rise to Kac modules for $m<0$
   and anti Kac modules for $m>0$. The remaining modules cannot
   be obtained through Verma modules of the sort considered
   there.} This statement is particularly clear for $m=0$ when
   the singular vector is a ground state. In view of
   eq.~\eqref{eq:SFKac} the attentive reader will have
   anticipated that the residual cases $e=mk$ simply arise by
   applying the spectral flow automorphism $\gamma_m$.
\smallskip

   The structure of the Kac modules may be inferred from
   their composition series. According to our previous statements
   the Kac module $\langle mk,n\rangle$ contains precisely one
   irreducible submodule denoted by $\langle n-1\rangle^{(m)}$. The
   quotient of $\langle mk,n\rangle$ by the submodule
   $\langle n-1\rangle^{(m)}$ turns out to be the irreducible
   representation $\bigl(\langle n\rangle^{(m)}\bigr)'$. Hence,
   one can describe the representation using the composition series
\begin{equation}
   \label{eq:CompS}
   \begin{split}
     \langle mk,n\rangle:&\quad \bigl(\langle n\rangle^{(m)}\bigr)'
     \ \longrightarrow \ \langle n-1\rangle^{(m)}\ \ .
   \end{split}
\end{equation}
   Again, all this can be understood best for $m=0$ where the statement
   reduces to well-known facts about Kac modules of the finite dimensional
   subalgebra \gl. This remark especially implies that the atypical irreducible
   representations $\langle n\rangle^{(0)}$ are built over the
   one-dimensional \gl-module $\langle n\rangle$. They are transformed into
   the remaining representations $\langle n\rangle^{(m)}$ under the
   spectral flow automorphism $\gamma_m$. For $m\neq0$, the ground states
   of $\langle n\rangle^{(m)}$ can easily be seen to form the \gl-module
   $\langle mk,n-m\rangle$.
   The information contained in the composition series \eqref{eq:CompS} may
   be used to calculate the super-characters of the atypical irreducible
   representations $\langle n\rangle^{(m)}$. Following the ideas of
   \cite{Rozansky:1992td} one simply finds
\begin{equation}
   \label{eq:CharAtyp}
   \begin{split}
     \hat{\chi}_{\langle n\rangle}^{(m)}(\mu,\tau)
     &\ =\ \sum_{l=0}^\infty\hat{\chi}_{\langle mk,n+l+1\rangle}(\mu,\tau)\\[2mm]
     &\ =\ \frac{u^{n}}{1-uq^m}\,\frac{q^{\frac{m}{2}(2n+m+1)+1/8}
         \theta\Bigl(\mu-\frac{1}{2}(\tau+1),\tau\Bigr)}{\eta(\tau)^3}\ \ .
   \end{split}
\end{equation}
   Analogous results hold for anti Kac modules.
\smallskip

   Finally we need to discuss the projective covers of irreducible
   representations. The typical representations $\langle e,n\rangle$ with
   $e\not\in k\mathbb{Z}$ are projective themselves. But the atypical
   representations $\langle n\rangle^{(m)}$ have more complicated
   projective covers whose composition series reads
\begin{equation}
   \mathcal{P}_n^{(m)}:\quad
   \bigl(\langle n\rangle^{(m)}\bigr)' \ \longrightarrow \ %
   \langle n+1\rangle^{(m)} \oplus \langle n-1\rangle^{(m)}
   \ \longrightarrow \ \bigl(\langle n\rangle^{(m)}\bigr)'\ \ .
\end{equation}
   An alternative description of the projective covers is in terms
   of their Kac composition series
   $\mathcal{P}_n^{(m)}:\langle mk,n\rangle\to\langle mk,n+1\rangle'$.
   Consequently, the characters of projective covers are given by
\begin{equation}
   \hat{\chi}_{\mathcal{P}_n^{(m)}}(\mu,\tau)
   \ = \ \hat{\chi}_{\langle
mk,n\rangle}(\mu,\tau)-\hat{\chi}_{\langle
mk,n+1\rangle}(\mu,\tau) \ .
\end{equation}
   These statements can once again be checked explicitly for $m=0$
   and then generalized to arbitrary values of $m$ by means of the
   spectral flow transformation. For future convenience we shall
   silently omit the superscript $^{(m)}$ in the case that $m=0$.

\subsection{\label{sc:Mods}Some modular transformations}

   In this section we list the modular
transformations of all the affine characters
   appearing in the previous section. Since all these representations may be
   expressed in terms of Kac modules it is
sufficient to know the transformation
\begin{equation}
   \hat{\chi}_{\langle e',n'\rangle }(\mu,\tau)\ =\
   -\frac{1}{k}\int dedn\ \exp\frac{2\pi
i}{k}\Bigl[e'(n-1/2)+e(n'-1/2)+e'e/k\Bigr]\,
   \hat{\chi}_{\langle e,n\rangle }(\tilde{\mu},\tilde{\tau}) \ \ .
\end{equation}
   to derive the remaining ones. Using the series representation
   \eqref{eq:CharAtyp} one, e.g., obtains the following behavior
   for characters of atypical representations,
\begin{equation}
   \hat{\chi}_{\langle n'\rangle}^{(m) }(\mu,\tau)\ =\
   \frac{1}{2ki}\int dedn\ \frac{\exp2\pi
i\bigl[e/k(n'+m)+m(n-1/2)\bigr]}{\sin(\pi e/k)}\,
   \hat{\chi}_{\langle e,n\rangle }(\tilde{\mu},\tilde{\tau})\ \ .
\end{equation}
   Similarly, using the Kac composition series for projective
   covers we deduce
\begin{equation}
   \begin{split}
     \hat{\chi}_{\mathcal{P}_{n'}^{(m)}}(\mu,\tau)
     &\ =\ \hat{\chi}_{\langle mk,n'\rangle
}(\mu,\tau)-\hat{\chi}_{\langle mk,n'+1\rangle }(\mu,\tau)\\
     &\ =\ \frac{2i(-1)^m}{k}\int dedn\ \exp2\pi i\Bigl[e/k(n'+mk)+mn\bigr]\,
          \sin(\pi e/k)\,\hat{\chi}_{\langle
e,n\rangle }(\tilde{\mu},\tilde{\tau})\ \ .
   \end{split}
   \raisetag{48pt}
\end{equation}
   The alternating signs in these formulas arise since the spectral flow
   changes the parity of representations for odd values of $m$.

\subsection{\label{sc:AffFus}Fusion rules of the \agl\ current algebra}

   Up to the need to incorporate the spectral flow automorphism and
   the additional atypical representations induced from it, the fusion
   rules of \agl\ agree precisely with the tensor product decomposition
   of \gl-modules, see e.g.\ \cite{Gotz:2005jz}. Given any two integers,
   $m_1, m_2 \in \mathbb{Z}$, we thus find
\begin{align}
     \langle e_1,n_1 \rangle &\otimes \langle e_2,n_2 \rangle \!\!\!\!\!&\cong\ &%
     \begin{cases}
       \langle e_1+e_2,n_1+n_2\rangle' \oplus \langle e_1+e_2,n_1+n_2-1\rangle
         &,\ e_1{+}e_2\not\in k\mathbb{Z}\\[2mm]
       \mathcal{P}^{(m)}_{n_1+n_2-1}
         &,\ e_1{+}e_2=mk
     \end{cases}\nonumber\\[2mm]
     \langle n_1 \rangle^{(m_1)}&\otimes \langle
n_2 \rangle^{(m_2)} \!\!\!\!\!&\cong\ &\langle n_1+n_2 \rangle^{(m_1+m_2)} \nonumber \\[2mm]
     \langle n_1 \rangle^{(m_1)}&\otimes \langle
e_2,n_2 \rangle \!\!\!\!\!&\!\!\!\!\cong\ &\langle m_1k+e_2, n_1+n_2 \rangle\ \ .
   \raisetag{16pt}
\end{align}
   The prime $'$ in the first line indicates that the representation
   has the opposite parity compared to our standard choice.

\def\cprime{$'$} \def\cprime{$'$}
\providecommand{\href}[2]{#2}\begingroup\raggedright\endgroup

\end{document}